# Cumulants, Moments and Selection

Hasan Ahmed*[1], Deena Goodgold*[2], Khushali Kothari*

*Authors listed in arbitrary order

[1]Department of Biology, Emory University, Atlanta, GA 30322, USA. [2]Department of Biological Sciences, University of Idaho, Moscow, ID 83844, USA.

## Abstract

We first describe a fundamental connection between cumulants/moments and selection — which follows intuitively when heterogeneity is added to Matlthus's population model. In doing so we provide an intuitive explanation of cumulants widely but incorrectly regarded as having no such interpretation. These fundamental relations are more general than Fisher's fundamental theorem of natural selection — allowing for calculation of the standard deviation, skewness and kurtosis of fitness far into the future — and are also more precise; indeed it becomes clear that Fisher's theorem is incorrect for fitness in the conventional/natural sense. Thanks to the close connection between selection and moments, a simple relation also exists between the moments of fitness and the moments of mutation — at equilibrium and also over time; many biologically meaningful claims follow as logical consequences with connections to Haldane's load theory and a more general formula for coefficient of variation of fitness.

## Significance statement

The goal of our paper is to examine in a systematic manner the significance of a very close connection between cumulants and moments and continuous and discrete selection respectively. This connection has several significant consequences. It provides 1) an intuitive definition for cumulants and moments, 2) a complementary pair of equations which are more general and more precise than Fisher's fundamental theorem of natural selection and 3) justification for defining fitness in certain ways. It also creates 4) a simple order between the moments of fitness and the moments of mutation. And notably 5) many results in theoretical biology are special cases or consequences of this simple order.

## 1. Introduction

Cumulants and moments are key concepts in statistics; selection is a fundamental concept in evolutionary biology and has application to a variety of topics from other fields such as depletion of susceptibles in epidemiology [1], economic development [2] and waning of immune memory [3]. It turns out that cumulants and moments are very closely related to selection. In particular, cumulants correspond to the basic mathematics of continuous growth/selection, and moments correspond to the basic mathematics of growth/selection over discrete intervals.

Although a moment is simply the expected value of a random variable raised to a (nonnegative) integer power, cumulants are often regarded as having no simple interpretation [4]. In fact there is a simple interpretation of cumulants in the context of evolution. This fact was noted by [5] as early as 1992 in a paper that was largely ignored. Later expanded upon by [6], it is now known to theoretical evolutionary biologists [7] [8], but it is hardly known outside that field. Hence one of the goals of this paper is to explain this relationship between cumulants and continuous selection to a wider group of people (§ 1.1) while also examining its limitations (§ 3-5).

In practice, however, populations do not grow continuously; they grow in discrete steps (births, deaths, cell division). So instead of the instantaneous growth rate ($r$), it is also possible to think about the fold expansion ($R$) over a unit of time.

$$R = e^r \quad (1)$$

If variation in generation interval by genotype is not too great, it can be convenient to define $R$ as the fold expansion over a single generation (or an integer number of generations) — see § 4.

The goal of this paper is primarily conceptual. Nonetheless we also provide an easy to use quantitative framework for assessing the effects of deleterious mutation and long-term selection relative to other factors such as short-term selection. One of our aims is to formulate theory which contributes to the understanding of biologists and scientists more generally outside of theoretical biology. Experts in theoretical evolutionary biology will be familiar with some of our specific claims but not necessarily their derivation in this context. For such readers the relative advantages of Eq 4 vs Eq 5A/5B and the moment formulas in the context of mutation are key points to note. We do not intend this paper to be pedagogic, but we do raise issues that are very relevant for pedagogy.

A brief outline of our paper follows. 1) By generalizing Malthus's population model, we provide an intuitive definition for cumulants and moments in § 1.1. 2) A complementary pair of equations for the effect of selection on the fitness distribution -- Eq 4 and Eq 5A/5B -- follow intuitively from that definition. We show the relative advantages and disadvantages of these equation (§ 4, § 5.2). The previously noted [6] dynamical insufficiency of Eq 4 becomes largely a moot point. 3) For a generic deleterious mutation model, we provide a single equation -- Eq 9 -- which provides *all* of the moments of fitness at equilibrium and a similar equation for the moments over time in Supplement § 5. 4) Certain biologically meaningful claims follow as logical consequences or special cases of Eq 9 namely A) scale and shape parameters for the equilibrium fitness distribution (§ 5.3), B) the mean fitness formula from Haldane's load theory is reproduced but in a clearer context (§ 5.3, § 6), C) a necessary and sufficient condition for mutation selection balance (§ 5.3), D) an important but subtle change from recombination (§ 5.3.2) and E) a more general formula for coefficient of variation in fitness (§ 5.3.3). An important overarching point is the very close link between mathematical theory and scientific concepts mediated by very simple intuitive dynamical systems.

Incidentally it also becomes clear that selection is not fundamentally about over population even though that may have been an interpretation of Malthus's population model which guided evolutionary biology in its early years [9].

Mutation also provides interpretation to certain moment sequences in particular those moment sequences that correspond to mutation-selection balance.

*Basic assumptions.* In this paper we focus on intrinsic fitness ($r$, $R$) rather than realized fitness ($r^+$, $R^+$). Although we allow for stochasticity at the individual level, we ignore population level stochasticity. "log" and "ln" both refer to natural log.

## 1.1. Cumulants, moments and population growth

According to Malthus's famous population model, if birth and death rates are constant, a population will grow in a "geometric" manner [10] i.e. the ratio of the current population relative to one unit of time in the past will remain constant over time. In this case the fold increase over time is given by the simple equation -- $R^t$. It will soon become clear that this simple formula corresponds to the moments of $R$ in the degenerate fully homogenous case.

An obvious extension of Malthus's model is to consider an ensemble of such populations each with a potentially different growth rate. In this context, heterogeneity in growth rates is precisely what is meant by selection. And the *fold change* over 1, 2 or 3 units of time is exactly equal to the 1st, 2nd or 3rd moments of $R$ (respectively) of the ensemble population. As indicated by Eq 2 this same pattern holds for all higher moments as well.

$$\frac{P(t)}{P(0)} = M_t(R) \text{ for t in } [0, 1, 2, \ldots] \tag{2}$$

where $P(0)$ is the current population, $P(t)$ is the population at time $t$ and $M_t(R)$ is the $t$-th moment of $R$.

In a more vague sense the cumulants of $R$ also have an interpretation in this context as a surprise element in population growth when the population at previous time steps is known. However, this interpretation of cumulants as residuals is precise only under a very peculiar predictive model and only valid for nonnegative variables. A much simpler and more general interpretation exists if we instead think in terms of exponential growth rate ($r$). In this context the cumulant generating function of $r$ ($CGF_r$) is exactly equal to the *natural log of the fold change* in population.

$$CGF_r(t) = \ln\left(\frac{P(t)}{P(0)}\right) \quad (3)$$

Since cumulants are the derivatives of this function, it follows that the 1st cumulant of *r* is the 1st derivative of log population, the 2nd cumulant of *r* is the 2nd derivative of log population and so on. This provides a simple enough interpretation of cumulants in the context of population growth. The key to understanding the next section is to realize that *the 1st derivative of log population is also the mean of r in the population*; Eq 4 follows logically from this fact.

## 1.2. *r* and cumulants

In the simple scenario where long term selection is the only force (so every lineage in the population has a specific growth rate), a straightforward and exact relationship exists between the cumulants of a fitness distribution as they evolve over time.

$$\frac{dK_a(r)}{dt} = K_{a+1}(r) \quad (4)$$

where $K_a(r)$ is the *a*-th cumulant of the fitness distribution if fitness is measured in terms of the Malthusian parameter (*r*). So the 2nd cumulant (variance) of *r* is the 2nd derivative of log population and the 1st derivative of the mean of *r*; indeed the fact that variance must be nonnegative corresponds to the increase of fitness under selection. The 3rd cumulant of *r* is the 3rd derivative of log population, the 2nd derivative of the mean of *r* and the 1st derivative of the variance of *r*. And this pattern continues for higher cumulants. Notably, although we typically think of selection as reducing variability, the opposite result in which the variance of *r* increases because of selection is also possible and corresponds to a positive 3rd cumulant (and hence also positive skewness) of log fitness.

Note that: because the 3rd and 4th cumulants are closely related to skewness and excess kurtosis, we some times refer to them respectively as "unscaled skewness" and "unscaled excess kurtosis".

Although Eq 4 follows intuitively from the analogy with population growth, for a simple algebraic derivation see [5].

Although it is possible to construct exceptions, low vs high variance, negative vs positive skewness and negative vs positive excess kurtosis -- all have a characteristic look as shown by the solid lines in Figure 1. Looking at the probability density functions, high variance tends to be spread out further from the mean, positive skewness tends to decline more slowly towards the right (and have mean>median), and positive excess kurtosis tends to have fatter tails. We also show how these probability distributions evolve under selection (dashed lines in Figure 1). Note that the negative skewness and positive skewness distributions in the middle panel have the same variance initially but over time the positive skewness distribution becomes noticeably much more spread out. The negative excess kurtosis and positive excess kurtosis distributions in the bottom panel both have skewness=0 initially, but over time the former develops the characteristic look of negative skewness, and the latter develops the characteristic look of positive skewness.

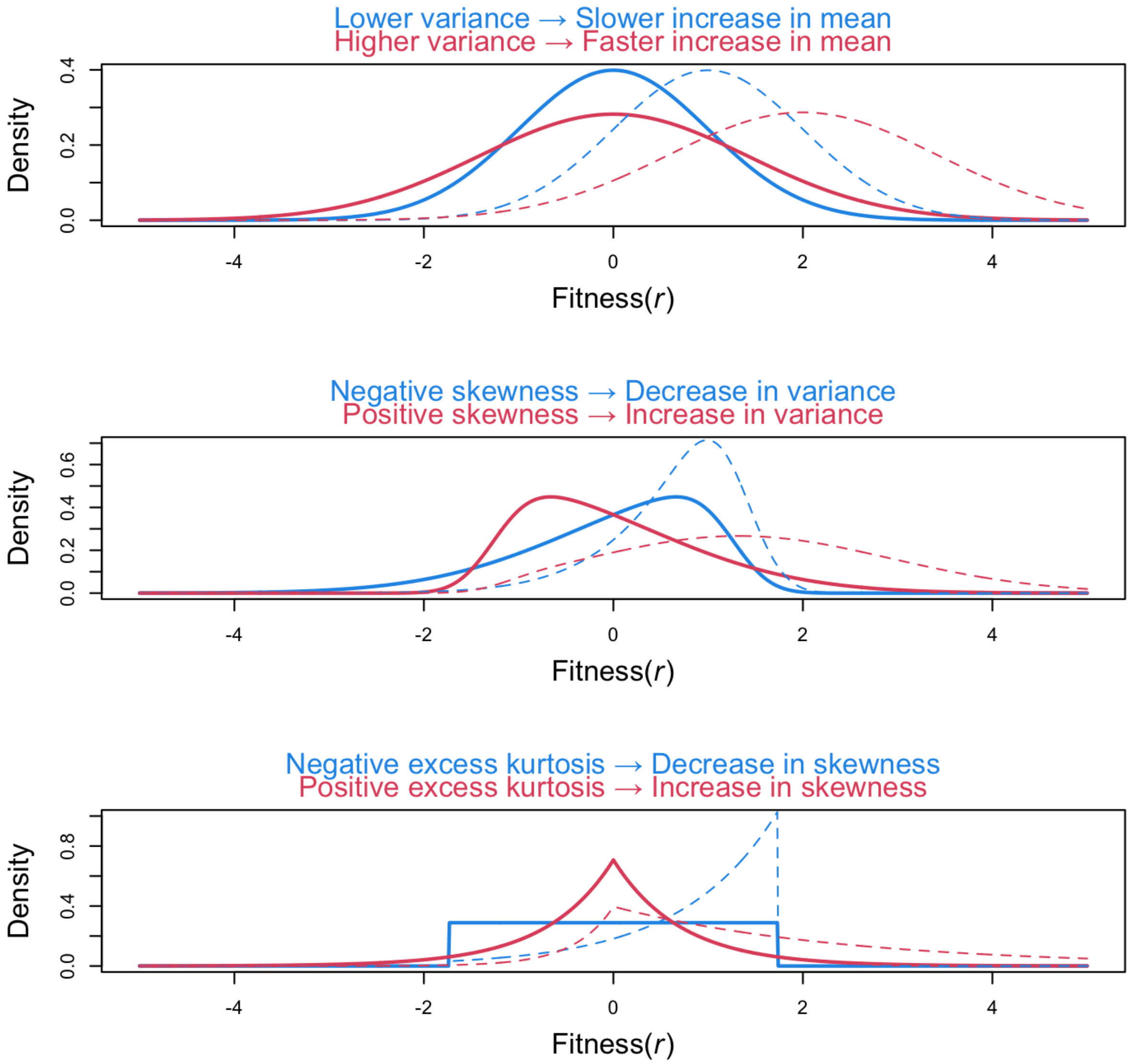

*Figure 1. Understanding cumulants and selection visually. In this figure the top plot compares high variance (red solid line) and low variance (blue solid). The middle plot compares positive skewness (red solid line) and negative skewness (blue solid line). The bottom plot compares negative excess kurtosis (blue solid line) and positive excess kurtosis (red solid line). The dashed lines show how the fitness distributions evolve after selection over one unit of time.*

# 2. *R* and moments

In addition to looking at the continuous growth rate ($r$), we can also consider the fold increase ($R$) of a lineage over a unit of time; $R=e^r$. Again we consider a situation where selection is the only force (no genetic drift, no mutation, no recombination, constant environment). If we now quantify fitness in terms of $R$, there is an exact and straightforward formula for how the raw moments of the fitness distribution change over time.

$$M_{a,t+1}(R) = \frac{M_{a+1,t}(R)}{M_{1,t}(R)} \tag{5A}$$

where $M_{a,t+1}(R)$ is the value of $a$-th moment at time=$t+1$. This formula applies also to fractional and negative 'moments'. More generally,

$$M_{a,t+\Delta t}(R) = \frac{M_{a+\Delta t,t}(R)}{M_{\Delta t,t}(R)} \tag{5B}$$

So, instead of running an agent based simulation, Eq 5B can be used to calculate any moment of $R$ at any time point. Because of the mathematical and statistical properties of moments, the significance of knowing the moments is much greater than it may at first seem. In addition to the mean (1st moment), other well known statistical descriptives like coefficient of variation, skewness and excess kurtosis can be calculated from the moments using established formulas e.g. coefficient of variation = $M_2(R)/M_1(R)-M_1(R)$.

Once again, the reason for this identity (Eq 5B) becomes clear if we think in terms of population growth. The $a$-th moment is the fold expansion over $a$ units of time, so all we need to do to evolve the moments into the future is *to define subsequent fold expansion relative to that future population*. For algebraic derivation see Supplement § 1.

# 3. Fisher's fundamental theorem of natural selection

According to Fisher's fundamental theorem of natural selection, the rate of increase in mean fitness is the variance in fitness. Eq 4 and Eq 5A/5B generalize Fisher's fundamental theorem of natural selection to higher cumulants/moments, and they are also more precise. Indeed it becomes clear that Fisher's theorem is incorrect for fitness in the most conventional and natural sense.

If variation in generation interval can reasonably be ignored, then $R$ can be defined as fold expansion over a single generation. In this case the discrete change in $R$ over a single generation is a natural way to measure change in fitness which matches the definition of $R$ and the underlying biological process. From Eq 5A we can see that the discrete change in the first moment (i.e. mean) of $R$ is $M_2(R)/M_1(R)- M_1(R)$, which using the well known identity $M_2(R)=M_1(R)^2+Var(R)$, can be rewritten as $Var(R)/M_1(R)$. So Fisher's theorem is not valid under the most conventional/natural interpretation, but it is correct under certain alternative interpretations. In particular if fitness is quantified in terms of $r$ -- which is some times called "log fitness" to distinguish it from fitness in a more proper sense -- the *instantaneous* change in the mean is the variance; this corresponds to Eq 4 with *a=1*. Indeed Fisher was thinking in terms of *r* which he refers to as the "Malthusian parameter" [11].

# 4. *r* vs *R*

Although Eq 4 is more intuitive than Eq 5A/5B, there are advantages to the latter. Eq 5A/5B easily provide future values, whereas Eq 4 instead gives instantaneous rates of change. Secondly, the *r* scale gives disproportionate influence to low vs very low levels of fitness with *r* diverging to -∞ as $R \rightarrow 0$. Finally, if we cheat a bit and define *R* over generations instead of over time, we have a definition of fitness that more closely matches the biology and hence could be considered a more natural definition of fitness. This last advantage is illustrated by the following section on heterozygote advantage and again shows up in § 5.

## 4.1. Heterozygote advantage

In this section we assume that variability in generation interval by genotype is negligible and hence we define *R* as the fold expansion over one (sexual) generation. In this case *R* can have substantial advantages over *r* as illustrated by the following example. We consider a gene with two variants (A & B) where *R*=1 in homozygous AA individuals, *R*=1.2 in heterozygous individuals and *R*=0.01 in homozygous BB individuals along with random mating. At equilibrium, using the standard text book formula [12], *f* = (1.2-1)/(0.2+1.19) = 0.1439 where *f* is the frequency of variant B. We now consider fitness from the perspective of the gene variants. At equilibrium the *R* values for the two genes are equal.

$$R_A = f \cdot 1.2 + (1-f) \cdot (1) = 1.0288 \quad (6A)$$

$$R_B = f \cdot 0.01 + (1-f) \cdot (1.2) = 1.0288 \quad (6B)$$

However, the values for *r* are quite different.

$$r_A = f \cdot (\ln(1.2)) + (1-f) \cdot (\ln(1)) = 0.0262 \quad (6C)$$

$$r_B = f \cdot (\ln(0.01)) + (1-f) \cdot (\ln(1.2)) = -0.5066 \quad (6D)$$

There are problems on two levels. 1) Conceptually the *r* model assumes a constant growth rate for the 3 subpopulations (AA, BB & AB); this is unbiological unless there are many asexual generations between recombination events. 2) Even in that case, the *r* values given by Eq 6C and Eq 6D correspond to the instantaneous growth rate immediately following recombination, and the fact that the gene frequencies are at equilibrium is obscured. Defining *R* over 1 (sexual) generation avoids both these problems.

# 5. Mutation and selection

Here we consider a simple but generic model with deleterious mutations along with selection. The fitness of the child (*r*) is determined probabilistically by the following equation:

$$r = r^* - x \cdot y \quad (7A)$$

where $r^*$ is the fitness of the parent, $x$ is a random binomial variable that determines whether or not there is a deleterious mutation and $y$ gives the effect size of the mutation. The equivalent equation for $R$ is given by the following formula.

$$R = R^* \cdot e^{-x \cdot y} \tag{7B}$$

Although certain forms of epistasis can simply be subsumed into the probability distributions of $x$ and $y$ or have negligible effect, for conceptual clarity we ignore epistasis.

## 5.1. Mutation simulation for influenza

We first modeled the fitness distribution of a population of individuals over 4000 generations with a starting population size of 1 million. Each individual in the first generation had a fitness value of 1. Mutations occurred for each individual with a probability of 0.2 -- based on the mutation rate of influenza [13] [14]. The effects of mutations were determined by a gamma distribution for $y$ with shape=1 and rate=2.85 -- again based on influenza [15]. The probability of an individual reproducing was equal to their fitness value. For this reason, the model had some drift but was under dispersed. To maintain the population size, we doubled the population when it was less than 500,000 individuals.

We saw that the mean, variance, unscaled skewness and unscaled excess kurtosis approached and fluctuated around equilibrium values (Figure 2). The equilibrium values were essentially unchanged when the initial fitness distribution was changed to a discrete uniform over {0.1, 0.2, … 0.9, 1} though the initial change was then much more dramatic (Supplement § 2).

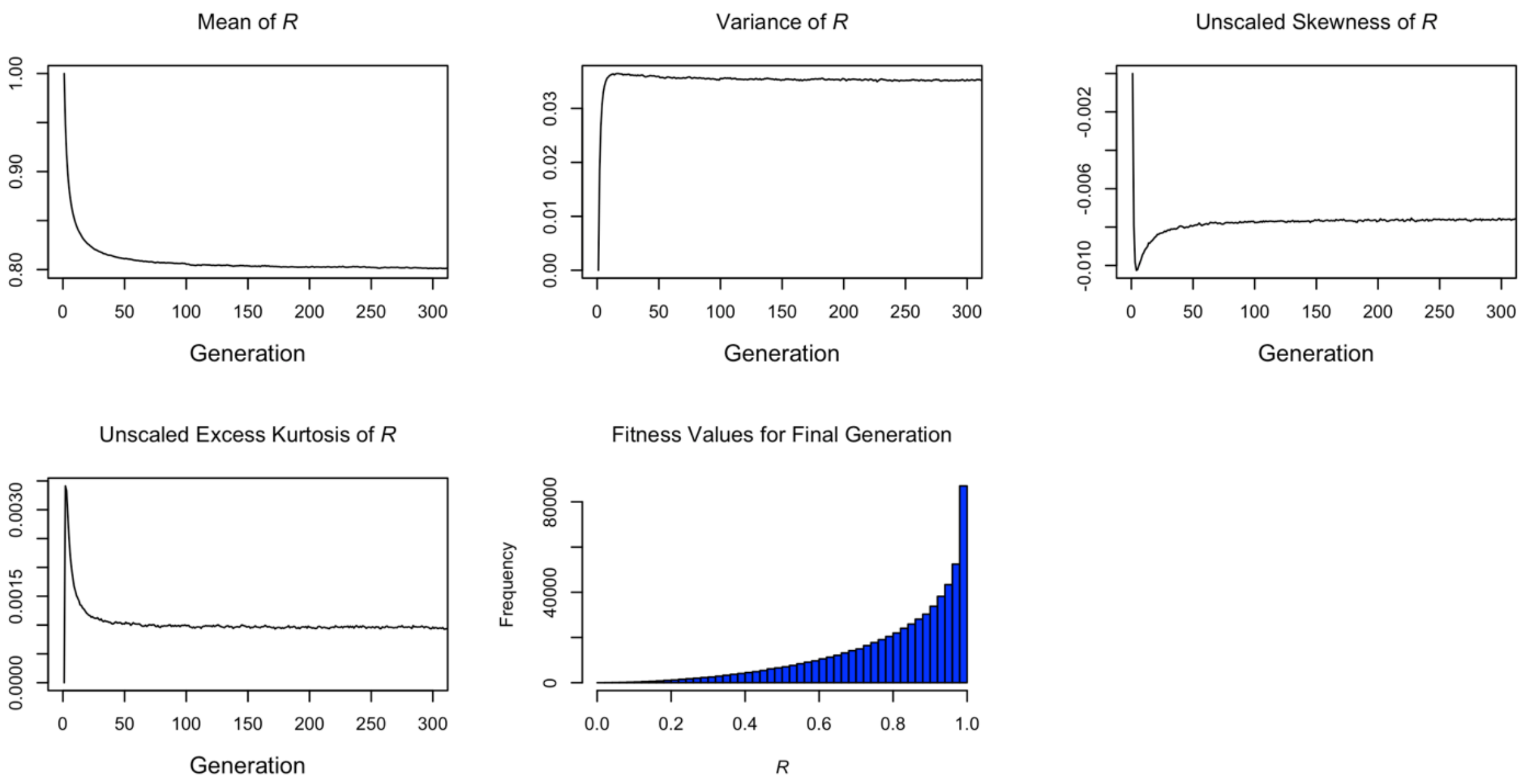

*Figure 2. Simulation results. All individuals in the initial generation have fitness of 1.*

## 5.2. Simulation versus theory — *r*

Because selection increases the mean of *r* by its variance and mutation decreases the mean of *r* by the mean of the mutation effect *(-x·y)* and likewise for the higher cumulants, we might think that at equilibrium:

$$\frac{K_{a+1}(r)}{K_a(-x{\cdot}y)} \approx -1 \quad (??) \tag{8}$$

where $K_a(r)$ is the *a*-th cumulant of the fitness distribution in terms of *r* and $K_a(-x{\cdot}y)$ is the *a*-th cumulant of $-x{\cdot}y$. But we see that this only very roughly holds (Table 1).

*Table 1. Average Cumulant Values for Generations 2000 to 4000 for r.*

| | *r* | *-x · y* |
|---|---|---|
| Mean | -0.262 | -0.0702 |
| Variance | 0.0957 | 0.0443 |
| Unscaled Skewness | -0.0697 | -0.0422 |
| Unscaled Excess Kurtoses | 0.0757 | |

*According to Eq 8 the values in red should be similar in magnitude but opposite in sign. Likewise for the values in blue and in green. But we see that this is not the case.*

In other words thinking of mutation and selection in a continuous way [6] does not very closely match the more biological discrete case. As expected, breaking each generation into multiple mini generations reduced the discrepancy between the simulation results and Eq 8 (Supplement § 3). This suggests that Eq 8 may closely hold in certain situations but not under those reported for influenza [13] [14] [15] -- nor for E. coli (Supplement § 4).

## 5.3. Simulation versus theory — *R*

If we think of selection in terms of *R* and the mutation model given by Eq 7A/7B, then we get a surprisingly simple equation (Eq 9) which gives *all* of the moments of *R* at equilibrium.

$$M_a(R) = \frac{\max(R)^a \cdot p^a}{\prod_{b=0}^{b=a-1} M_b(e^{-x\cdot y})} \tag{9}$$

Here $M_a(R)$ is the *a*-th moment of *R*. *p* is the probability of escaping deleterious mutation (i.e. the probability that $x$=0). *max(R)* is the value of *R* for the fittest individual in the initial population. $M_b(e^{-x\cdot y})$ is the *b*-th moment of $e^{-x\cdot y}$, the multiplicative effect of mutation (equivalently the fitness of the child relative to its parent). Since *R* is bounded, its moments uniquely determine its probability distribution [16].

In addition to the equilibrium moments, it is also possible to calculate all the moments over time from which Eq 9 can be derived as the limit case (Supplement § 5). Notably that equation for moments over time also holds when there is a mix of advantageous and deleterious mutations. In that case fitness keeps on increasing and no equilibrium is reached.

Eq 9 connects to many important questions/topics in evolution and provides an alternative perspective. We see that an equilibrium fitness distribution exists and that it is determined solely by *max(R)* and the distribution of mutation effects. The former acts as a scale parameter, having the same effect on the moments as multiplication by a constant, whereas the shape is determined solely by mutation. Equilibrium mean fitness is simply $max(R)\cdot p$ -- consistent with Haldane's load theory [17]. And $p>0$ is the essential criteria for mutation selection balance i.e. at least a few offspring must escape deleterious mutation. If no offspring escape deleterious mutation ($p=0$), all of the moments, and hence the fitness distribution as a whole, will collapse towards 0.

In contrast to the previous section, the simulation results now match the theoretical predictions very closely (Table 2).

*Table 2. Average Cumulant Values for Generations 2000 to 4000 for R.*

| | *R* (simulations) | *R* (theoretical equilibrium) |
|---|---|---|
| Mean | 0.800 | 0.8 |
| Variance | 0.0351 | 0.0351 |
| Unscaled Skewness | -0.00757 | -0.00757 |
| Unscaled Excess Kurtoses | 0.000952 | 0.000951 |

*The theoretical equilibrium values for the moments were taken from Eq 9. These were converted to cumulants using the standard formulas [18].*

### 5.3.1. Special case when $-x\cdot y$ is compound Poisson

It is possible to formulate $x\cdot y$ in terms of the number of mutations and the effect per mutation: $x\cdot y = \Sigma z_j$ where the $z$s are the effects of the mutations. If the $z$s are independent and the number of mutations is Poisson distributed (i.e. $-x\cdot y$ is compound Poisson distributed) then:

$$M_b(e^{-x\cdot y}) = e^{\lambda\cdot(M_b(e^{-z})-1)} \quad (10)$$

where $\lambda$ is the mean number of de novo deleterious mutations per individual and $M_b(e^{-z})$ is the $b$-th moment for the effect on $R$ of a single deleterious mutation. The already surprisingly simple Eq 9 becomes even simpler under this formulation. We note for $\lambda=1$ and $z$~gamma(shape=$s$, rate=1) an especially elegant form for the equilibrium moments in terms of harmonic numbers: first moment of $e^\wedge(-1^{-s})$, second moment of $e^\wedge(-1^{-s}-2^{-s})$, third moment of $e^\wedge(-1^{-s}-2^{-s}-3^{-s})$ and so on. Because $CGF_r(a) = ln(M_a(R))$ cf

Eq 2 and Eq 3, in this case via analytic continuation it is also possible to calculate the cumulants of $r$; these also have an elegant form -- $K_a(r) = (-1)^a \cdot s^{(a)} \cdot \zeta(s+a)$ where $s^{(a)}$ denotes a rising factorial and $\zeta$ is the Riemann zeta function.

### 5.3.2. Recombination

Under a simple model of recombination (multiple segments with an independent probability of mutation which can reassort, random mating, no epistasis), Eq 9 stills applies but with one subtle but very important distinction. *max(R)* is now the maximum *R* that *could possibly exist under recombination*. The full significance of this will be discussed in a follow up paper. For now it should be noted that *max(R)* -- and hence the entire equilibrium distribution -- in this context is at least as high -- but potentially much higher -- than in the absence of recombination.

### 5.3.3. Coefficient of variation of *R*

From Eq 9, it is straightforward to derive the coefficient of variation (standard deviation divided by mean) of fitness.

$$CV(R) = \sqrt{M_1^{-1}(e^{-x \cdot y}) - 1} \qquad (11A)$$

where $M_1(e^{-x \cdot y})$ is the mean fitness (on the $R$ scale) of a child relative to its parent. Under the assumptions of section 5.3.1, a result of [19] is recovered as a special case.

$$CV(R) = \sqrt{e^{\lambda \cdot (1 - M_1(e^{-z}))} - 1} \qquad (11B)$$

where $\lambda$ is the mean number of de novo deleterious mutations per individual and $M_1(e^{-z})$ is the mean fitness (on the $R$ scale) of a child with a single deleterious mutation relative to its parent. Eq 10A is not only more general but also provides an alternative way to estimate the coefficient of variance. Table 3 illustrates the relationship between the coefficient of variation of $R$ and other key parameters.

*Table 4: Coefficient of variation of R and other parameters*

| Approximate Species | $\lambda$ | $M_1(e^{-z})$ | $M_1(e^{-x \cdot y})$ | $CV(R)$ |
|---|---|---|---|---|
| E. coli | 0.001 | 0.969 | 0.999969 | 0.6% |
| Humans | 2.1 | 0.991 | 0.981 | 13.8% |
| Influenza A | 0.223 | 0.761 | 0.948 | 23.4% |

*$\lambda$ is the mean number of de novo deleterious mutations per individual. $M_1(e^{-z})$ is the mean multiplicative effect on R of a single deleterious mutation. $M_1(e^{-x \cdot y})$ is the mean fitness of the child relative to its parent. CV is the coefficient of variation of R. Approximate species is a species that approximately matches the $\lambda$ and $M_1(e^{-z})$ values in the table according to the scientific literature. For $\lambda$ of 2.1 and $M_1(e^{-z})$ of 0.991 for humans see [20], [21]. The values for E. coli and Influenza A are based on our simulations described in supplement § 4 and main text § 5.1 respectively. The actual coefficient of variation for these species may be quite different because of factors such as short term selection.*

# 6. Discussion

Haldane's load theory has been badly misunderstood even by experts [22] [23]. To avoid such misunderstandings there are a few points we wish to stress. 1) If $p$ is substantially affected by the number of very mild mutations, the equations in Supplement § 5 provide the important context of how quickly equilibrium values are approached. 2) In the context of recombination, *max(R)* can be far larger than the fitness of the fittest individual [19], nonetheless the important quantity for determining the overall statistical distribution is *max(R)* as an *extrapolation relative to the bulk of the population* not the fitness of the fittest individual nor the fitness of a perfect individual. 3) We intended Eq 9 as a model for fitness variability caused by deleterious mutation and not necessarily as a prediction of overall variability.

We do not consider the cumulants and moments of traits other than fitness, a topic closely related to the Price equation, which we hope to address in the future. The Price equation is fully general because it uses realized fitness (*R+*) rather than intrinsic fitness (*R*) -- a distinction which is often elided. The Price equation is both a decomposition and a way to predict future values. In our view this has caused some unnecessary confusion. Eq 4 and Eq 5A/5B provide a middle ground between Fisher's fundamental theorem of natural selection and the Price equation, being more general and more precise than Fisher's fundamental theorem and making clear relations which are hidden under the generality of the Price equation. We view Eq 4 and Eq 5A/5B as complementary. The former gives an intuitive sense of the effects of selection (see Figure 1) while the latter is more useful for calculation.

For mutation, we do not consider diverging (advantageous mutation), collapsing ($p$=0), truly continuous and unbounded *max(R)* cases, exotic cases, which although unbiological may be of theoretical interest. Possibly interesting trajectories (Supplement § 5) and pairings (Eq 9) arise from mutation. And unlike the selection-only case, distinct trajectories can intersect. At minimum mutation gives us some elegant moment sequences (§ 5.3.1).

In addition to having a few experts understand theoretical issues well, it is also important that theoretical biology be formulated in a way that contributes to the understanding of scientists more generally. We provide an alternative perspective on core topics of evolutionary biology which may be helpful in this task. Ultimately we highlight an extremely close link between mathematical theory and scientific concepts which is mediated by intuitive and simple dynamical systems.

There are important caveats and limitations to our work. For conceptual clarity we assume $R$ is known. Nonetheless in naturalistic situations, where it can be difficult to disentangle $R$ from $R^+$, the fundamental relations provide a way to disentangle the moments of $R$ from the moments of $R^+$. We are not mathematicians, and we view the derivations in the supplement as informal derivations not rigorous mathematical proofs. We aim to be comprehensive and concise in explaining the key connections between cumulants/moments and evolutionary theory and hence we do not consider practical complications or dwell on the significance of any specific point to the extent that may be required in a pedagogic context. We do not consider short term selection, which we believe is a major factor. Rather we hope that the relations we have outlined can be useful in decomposing the relative effects of long term selection, deleterious mutation and short term selection.

# [Supplement] Cumulants, Moments and Selection

Hasan Ahmed, Deena Goodgold, Khushali Kothari

## 1. Derivation of discrete growth formula

We know that,

$$M_{a,t} = \int f_t(R) \cdot R^a \, dR$$

$$M_{a,t+\Delta t} = \int f_{t+\Delta t}(R) \cdot R^a \, dR$$

where $f_t$ is the probability density function for time = $t$ and $f_{t+\Delta t}$ is the probability density function at time = $t+\Delta t$.

By definition, a lineage with fitness of R grows $R^{\Delta t}$ fold over $\Delta t$ units of time, therefore

$$f_{t+\Delta t}(R) \propto f_t(R) \cdot R^{\Delta t} \rightarrow f_{t+\Delta t}(R) = c \cdot f_t(R) \cdot R^{\Delta t}$$

where $c$ is a normalizing constant.

$$\int c \cdot f_t(R) \cdot R^{\Delta t} \, dR = 1 \rightarrow c = \frac{1}{\int f_t(R) \cdot R^{\Delta t} \, dR}$$

Hence after substitution,

$$M_{a,t+\Delta t} = \int f_{t+\Delta t}(R) \cdot R^a \, dR = \int c \cdot f_t(R) \cdot R^{\Delta t} \cdot R^a \, dR = \frac{\int f_t(R) \cdot R^{a+\Delta t} \, dR}{\int f_t(R) \cdot R^{\Delta t} \, dR}$$

Here, we see that the numerator is equivalent to $M_{a+\Delta t,t}$ and the denominator is equivalent to $M_{\Delta t,t}$

Therefore,

$$M_{a,t+\Delta t} = \frac{M_{a+\Delta t,t}}{M_{\Delta t,t}}$$

## 2. Simulation results with initial discrete uniform distribution of *R* values

Figure S1 displays the cumulant values over generations 0 to 300 for a population with an initial discrete unform distribution of fitness values ($R$).

For this population the initial mean is 0.55 with the variance being higher since the initial fitness values range from 0.1 to 1 in a uniformly distributed manner. 10% of the population exists at each fitness value in {0.1, 0.2, ... 0.9, 1}. The initial unscaled skewness is 0 since the distribution is symmetric. The unscaled excess kurtosis is initially negative. High levels of selection occur early on in the simulation reflecting high initial variance, but the cumulant values plateau at a similar level compared to the simulation in the main text.

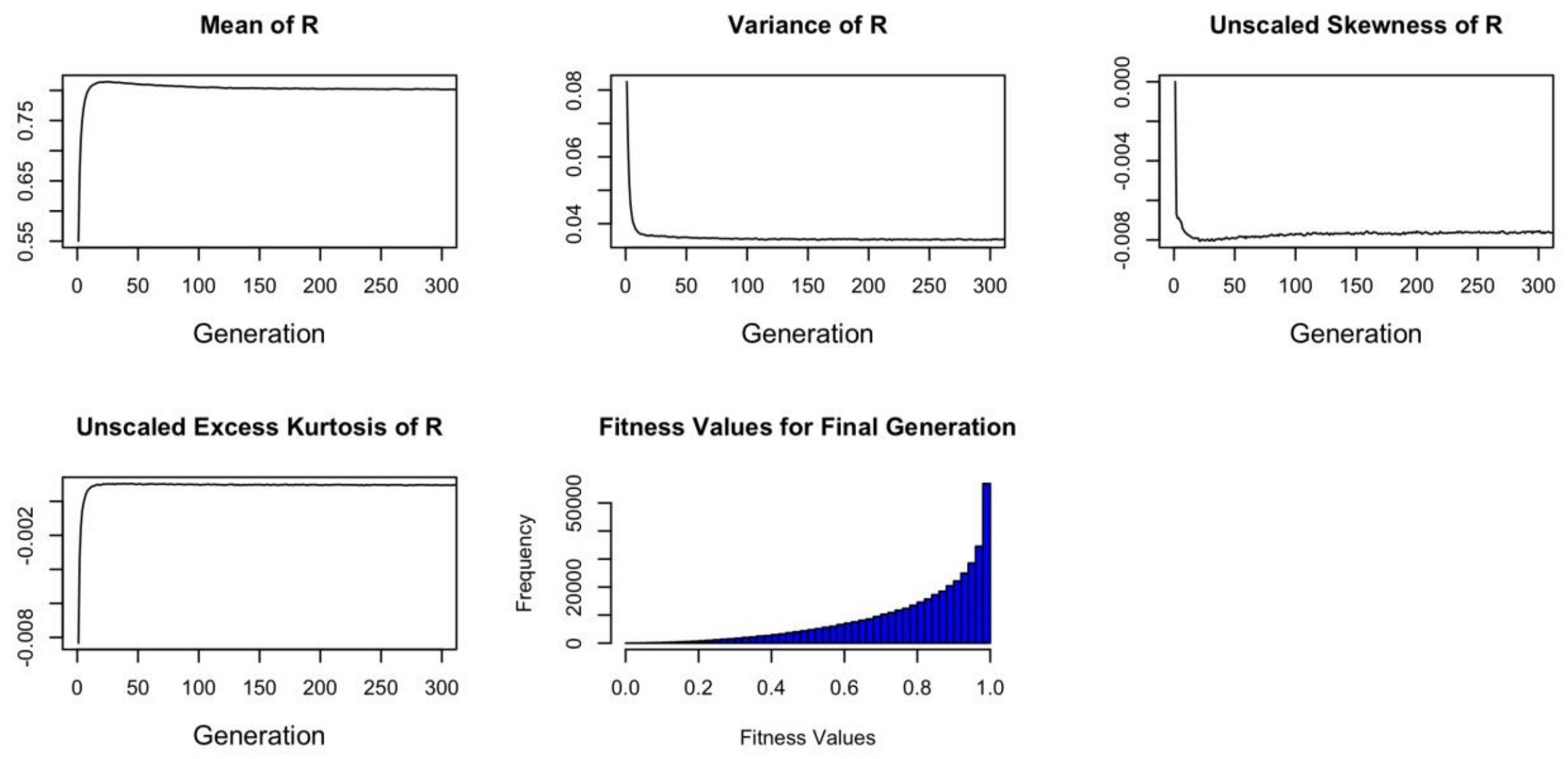

*Figure S1. Simulation results. Fitness values in the initial generation were distributed uniformly over {0.1, 0.2, ... 0.9, 1}.*

*Table S1. Average Cumulant Values for Generations 2000 to 4000 for r.*

| | r | $-x \cdot y$ |
|---|---|---|
| Mean | -0.261566 | -0.07017544 |
| Variance | 0.09575191 | 0.04432133 |
| Unscaled Skewness | -0.06968853 | -0.04216142 |
| Unscaled Excess Kurtoses | 0.07578221 | |

*Table S2. Average Cumulant Values for Generations 2000 to 4000 for R.*

| | *R* (simulations) | *R* (theoretical equilibrium) |
|---|---|---|
| Mean | 0.8002347 | 0.8 |
| Variance | 0.03509336 | 0.03506849 |
| Unscaled Skewness | -0.007572458 | -0.007565338 |
| Unscaled Excess Kurtoses | 0.0009511625 | 0.0009506762 |

These equilibrium cumulant values are essentially identical to the equilibrium cumulant values in the main text.

# 3. Simulation results with reduced mutation effect

Here we break one generation into 10 mini generations to make the simulation more like a continuous process. Mutations accumulate at each mini generation but at one tenth of the rate i.e. a 2% probability of mutation at each mini generation. Likewise, the amount of exponential growth over one mini generation is *r/10*. As expected, the simulation results much more closely conform to Eq 8 of the main text.

*Table S3. Average Cumulant Values for Generations 200 to 400 for r.*

| | $r$ | $-x \cdot y$ |
|---|---|---|
| Mean | -0.204 | -0.0702 |
| Variance | 0.0727 | 0.0443 |
| Unscaled Skewness | -0.0515 | -0.0422 |
| Unscaled Excess Kurtoses | 0.0546 | |

# 4. Mutation and selection in E. coli

Here, we run an analogous simulation to that in the main text except that the parameters for mutation were selected to match those reported for E. coli. Mutations occur for each individual with a probability of 0.001 based on the mutation rate of E. coli [1]. The effects of mutations are determined by a gamma distribution (with $\alpha = 3.03$ & $\beta = 194.24$ ) for 96.83% of mutations or by a standard exponential for the remaining 3.17% [2].

We see that the mean, variance, unscaled skewness, and unscaled excess kurtosis approach and fluctuate around an equilibrium value (Figure S2). The mean of R is approximately 0.9990018, the variance is low (3.07108e-05), the unscaled skewness is negative (-1.084431e-05, scaled skewness of -63.71845), and the unscaled excess kurtosis is positive (7.914932e-06, scaled excess kurtosis of 8391.989).

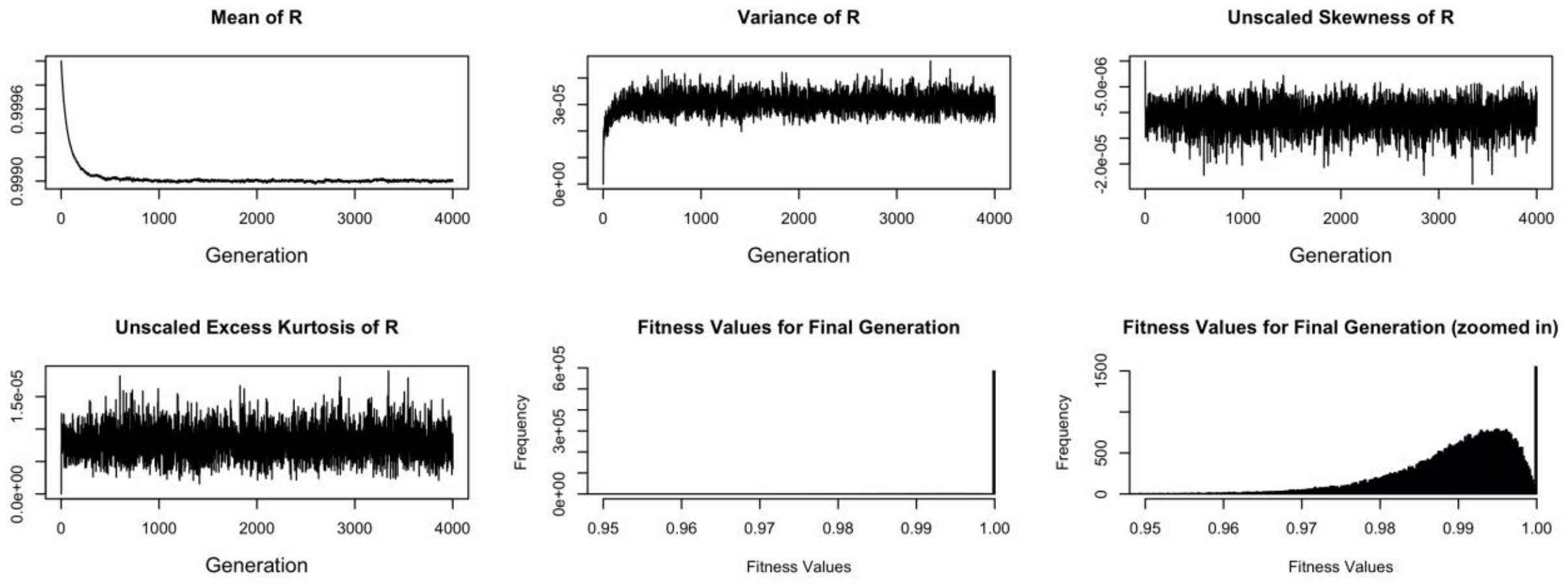

*Figure S2: Simulation results. All individuals in the initial generation have fitness of 1.*

*Table S4. Average Cumulant Values for Generations 2000 to 4000 for r.*

| | r | $-x \cdot y$ |
|---|---|---|
| Mean | -0.001026745 | -4.680476e-05 |
| Variance | 9.120267e-05 | 6.37112e-05 |
| Unscaled Skewness | -0.0002023346 | - 0.0001901992 |
| Unscaled Excess Kurtoses | 0.0007463787 | |

*Table S5. Average Cumulant Values for Generations 2000 to 4000 for R.*

| | *R* (simulations) | *R* (theoretical equilibrium) |
|---|---|---|
| Mean | 0.9990018 | 0.999 |
| Variance | 3.07108e-05 | 3.073879e-05 |
| Unscaled Skewness | -1.084431e-05 | -1.083727e-05 |
| Unscaled Excess Kurtoses | 7.914932e-06 | 7.898774e-06 |

# 5. Moments of *R* under mutation and selection

The formula for how moments of *R* change generation by generation is given by the following formula.

$$M_{a,g}(R) = \frac{M_{a+g,0}(R)}{M_{g,0}(R)} \cdot \frac{\prod_{b=a}^{a+g-1} M_b(e^{-x \cdot y})}{\prod_{b=0}^{g-1} M_b(e^{-x \cdot y})}$$

$M_{a,g}(R)$ is the *a*-th moment of fitness (*R*) for generation *g* with *g*=0 corresponding to the initial population, and $M_b(e^{-x \cdot y})$ is the *b*-th moment of the mutation effect function, specifically the moments of $e^{-x \cdot y}$ (Eq 7B in main text). We note that the selection effect and the mutation effect separate in a multiplicative manner. The above formula can be proved using induction.

For $g \geq a$, terms in the denominator and numerator of the second quotient cancel out, and the above formula can be more simply expressed as the following.

$$M_{a,g}(R) = \frac{M_{a+g,0}(R)}{M_{g,0}(R)} \cdot \frac{\prod_{b=g}^{a+g-1} M_b(e^{-x \cdot y})}{\prod_{b=0}^{a-1} M_b(e^{-x \cdot y})}$$

We now consider the behavior of this formula as *g* goes towards infinity. For finite *max(R)* the first quotient approaches $max(R)^a$, and the numerator of the second quotient approaches $p^a$ where *p* is the probability that there is *not* a deleterious mutation. This gives us Eq 9 in the main text.